\documentclass[12pt,preprint]{aastex}

\newcommand{\arcs}{\mbox{\ensuremath{^{\prime\prime}}}}
\newcommand{\arcm}{\mbox{\ensuremath{^{\prime}}}}

\usepackage{natbib}
\usepackage{longtable,lscape}
\begin{document}

\title{The Massive Star Forming Region Cygnus~OB2.\\ I.  {\it Chandra} catalogue of association members}

\author{N.J.~Wright and J.J.~Drake}
\affil{Harvard-Smithsonian Center for Astrophysics, 60 Garden Street, Cambridge, MA~02138}
\email{nwright@head.cfa.harvard.edu}

\begin{abstract}

We present a catalogue of 1696 X-ray sources detected in the massive star forming region (SFR) Cygnus~OB2 and extracted from two archival {\it Chandra} observations of the center of the region. 
A deep source extraction routine, exploiting the low background rates of {\it Chandra} observations was employed to maximize the number of sources extracted. 
Observations at other wavelengths were used to identify low count-rate sources and remove likely spurious sources. 
Monte Carlo simulations were also used to assess the authenticity of these sources. 
X-ray spectra were fitted with thermal plasma models to characterize the objects and X-ray light curves were analyzed to determine their variability. 
We used a Bayesian technique to identify optical or near-IR counterparts for 1501 (89\%) of our sources, using deep observations from the INT Photometric H$\alpha$ Survey, the Two Micron All Sky Survey, and the UKIRT Infrared Deep Sky Survey-Galactic plane Survey. 755 (45\%) of these objects have six-band $r'$, H$\alpha$, $i'$, $J$, $H$, and $K_s$ optical and near-IR photometry. 
From an analysis of the Poisson false-source probabilities for each source we estimate that our X-ray catalogue includes $<$~1\% of false sources, and an even lower fraction when only sources with optical or near-IR associations are considered. 
A Monte Carlo simulation of the Bayesian matching scheme allows this method to be compared to more simplified matching techniques and enables the various sources of error to be quantified. 
The catalogue of 1696 objects presented here includes X-ray broad band fluxes, model fits, and optical and near-IR photometry in what is one of the largest X-ray catalogue of a single SFR to date. The high number of stellar X-ray sources detected from relatively shallow observations confirms the status and importance of Cygnus~OB2 as one of our Galaxy's most massive SFRs.\\

\end{abstract}


\section{Introduction}

The vast majority of stars are believed to form in clusters with sizes ranging from the nearby small star forming regions (SFRs) typified by Orion and others in the Gould Belt \citep[e.g.,][]{hill00,popp97}, up to distant massive super star clusters such as W49 and Westerlund~1 \citep[e.g.,][]{clar05b}.  The latter are thought to be the major star factories in the Universe \citep[e.g.,][]{elme85,lada03}, each containing hundreds to thousands of OB stars and millions of low mass stars.  Understanding how these regions form and evolve is vital to our comprehension of the first stars and starburst galaxies, the factors controlling the stellar initial mass function \citep[IMF; e.g.,][]{salp55,scal86}, and the structure and evolution of our Galaxy.

Our appreciation of the formation of massive SFRs and our understanding of the influences shaping star and planet formation within them is still only fragmentary \citep[e.g.,][]{chab03}. Incisive studies of massive SFRs are hindered by their scarcity and great distances, resulting in both an inability to probe the low masses necessary for a complete grasp of the IMF, and insufficient spatial resolution to diagnose the physical processes at work. Accordingly it is important to make full use of the very few nearby opportunities we have to pursue these issues. Cygnus~OB2 is one such region at a distance of only $\sim$1.45~kpc \citep{hans03}: it is one of the most massive OB associations in our Galaxy and hosts a tremendously rich and diverse stellar population \citep[e.g.,][]{mass91,come02,vink08,negu08}. The first detailed study of the association was conducted by \citet{redd67} who found several hundred OB stars and estimated a total mass of 6000-27000~M$_{\odot}$, despite the high extinction towards the region ($A_V \sim 4-8$). As deeper observations have penetrated the extinction, the perceived size and observed morphology of the region have evolved. Using near-infrared photometry, \citet{knod00} estimated a total stellar mass for the association of $(4-10) \times 10^4$~M$_{\odot}$ and inferred the presence of 2600~OB stars. This has led to comparisons with globular clusters and the superclusters that characterize massive extragalactic SFRs. Doubts have been thrown on these conclusions by \citet{hans03}, who argued for a closer and less massive association, and \citet{drew08} who found evidence for a greater spread of ages.

Understanding massive SFRs and disentangling the complex structures requires a complete stellar census of association members. X-ray observations offer a largely unbiased marker of youth that is highly effective in separating young association members from older field stars. This is because pre-main sequence (PMS) stars are typically $10^1 - 10^4$ times more luminous in X-rays than main sequence stars \citep{prei05}, due to enhanced magnetic activity (for low mass stars) and collisions in strong stellar winds (for high mass stars). Studies of young SFRs using X-ray observations are also not biased toward high-mass sources (as are studies based in the optical) or sources with protoplanetary disks (for studies in the infrared). These advantages, combined with the ability to penetrate heavy extinction up to $A_V \sim 500$ \citep{gros05} and the sub-arcsecond resolution of the {\it Chandra X-ray Observatory} \citep{weis00} make X-ray observations an ideal method for studying SFRs in the Galactic plane.

In this paper we present a catalogue of point sources identified in the center of Cygnus~OB2 using {\it Chandra} archival observations of two fields in the region. One of these fields was studied by \citet{alba07}, who employed a conservative significance threshold to reduce spurious detections and combined the observations with the Two Micron All Sky Survey (2MASS) near-IR photometry to determine stellar properties for 519 X-ray sources. 
However, X-ray observations with low background rates enable the detection of sources with very few counts. Furthermore, even fainter sources with as few as 3 counts can be identified by using known source lists at other wavelengths to remove spurious detections. We pursue this approach in this paper.

In Section~2 we describe the archival X-ray observations (originally presented by \citealt{alba07} and \citealt{butt06}), perform source detection, photon extraction, and characterize the spectra and light curves of the identified sources. 
In Section~3 we cross-correlate this source list with data from three deep photometric surveys of the Galactic plane and recent spectroscopic studies of Cyg~OB2. 
In Section~4 we present the complete catalogue, including tables covering the extracted source properties, spectral model fits, and optical and near-IR associations. 
Finally, in Section~5 we perform a statistical analysis of the low-significance sources detected here to show that they are valid stellar X-ray members of Cyg~OB2.  
Future papers will discuss the properties of the identified stellar populations and their relation to the structure of Cygnus~OB2.

\section{{\it Chandra} X-ray observations and data reduction}

In this section we describe the process of reducing {\it Chandra} X-ray observations, detecting sources, and performing photon extraction. Two fields in the region of the Cyg~OB2 association have been observed with the Advanced CCD Imaging Spectrometer (ACIS) on board {\it Chandra} and were retrieved from the {\it Chandra} Data Archive\footnote{http://cxc.harvard.edu/cda/} (see Table~\ref{log} for details). Both observations utilize ACIS-I, which comprises four CCDs (chips I0-I3), each with $1024 \times 1024$~pixels (at a scale of 0$\farcs$.492~pix$^{-1}$) giving a $17\arcm \times 17\arcm$ field of view (FoV). Some of the ACIS-S chips were turned on during the observations, but due to the high off-axis angle to these chips and consequently large point spread function (PSF), we do not analyse these data here.

\begin{figure*}
\epsscale{1.0}
\plotone{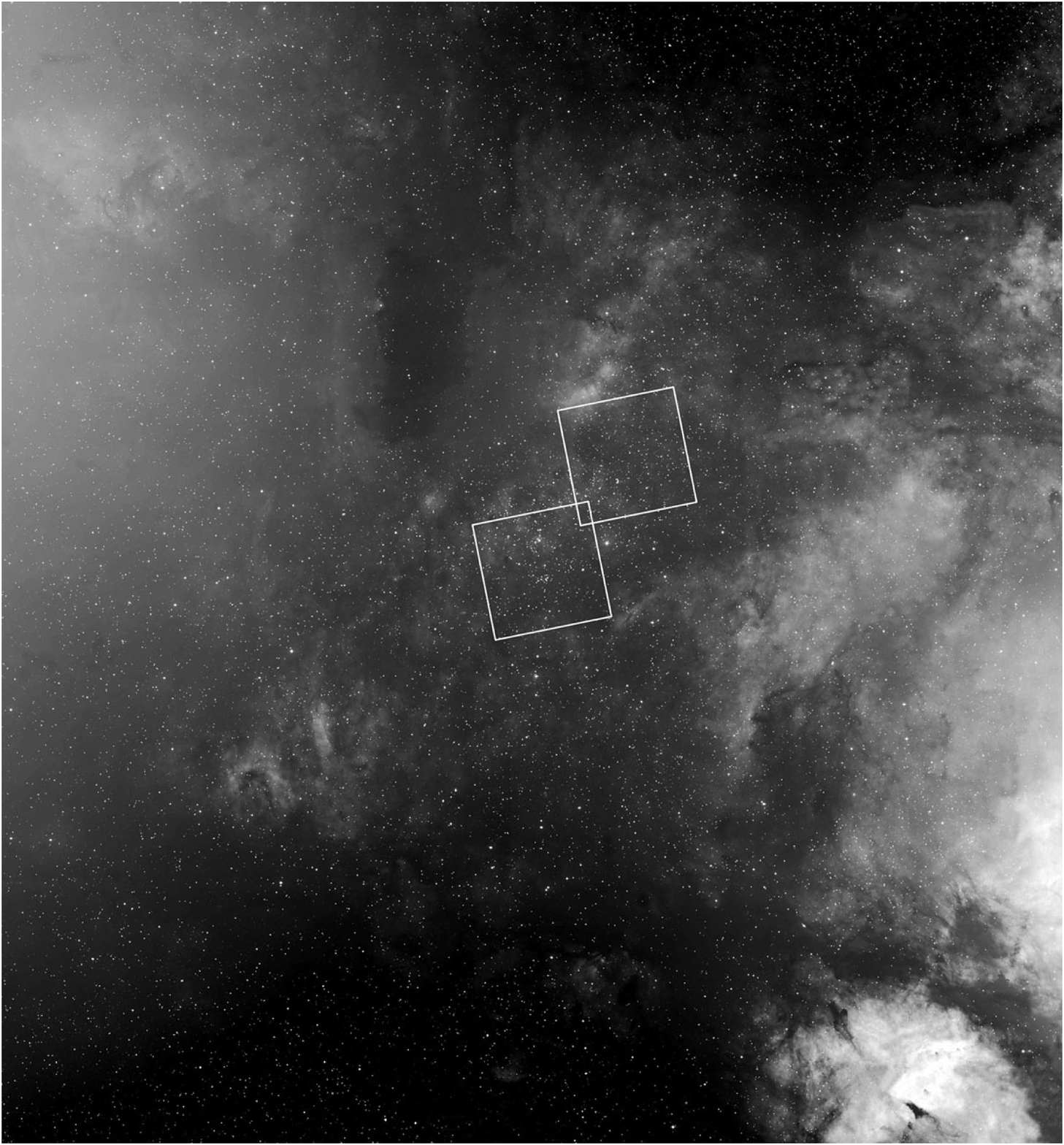}
\caption{H$\alpha$ image of the Cygnus OB2 region mosaiced from IPHAS observations and displayed using a logarithmic intensity scale. The image size is $2.67 \times 2.85$ degrees and centered on (RA, decl.) = (20:33:00, +41:15:00) with north up and east to the left. The $17\arcm \times 17\arcm$ fields of view of the two {\it Chandra}-ACIS observations discussed in this paper are shown as white squares.}
\label{cygnus_region}
\end{figure*}

The deeper of the two fields covers the central region of the association itself (pointing centered on RA~=~20:33:11.0 and decl.~=~+41:15:10.00) and was observed on 2004 January 16 (Observation ID~4511; PI:~E.~Flaccomio) with a total exposure time of 97.7~ks in {\sc very faint} mode. A second {\it Chandra} observation, north west of the Cyg~OB2 center (at RA~=~20:32:07.0 and decl.~=~+41:30:30.00) was observed on 2004 July 19 (Observation ID~4501; PI:~Y.~Butt) with an exposure time of 49.4~ks in {\sc very faint} mode. Designed to characterize the previously unidentified gamma-ray source TeV~J2032+4130 \citep{butt06} reported by the High Energy Gamma-Ray Astronomy collaboration, the stellar observations have not been studied in detail. Figure~\ref{cygnus_region} shows the positions of the two {\it Chandra} observations superimposed on an IPHAS \citep[INT Photometric H$\alpha$ Survey, see Section~\ref{s-cc},][]{drew05} H$\alpha$ mosaic of the region.

\subsection{Data reduction}

Reduction of the {\it Chandra} observations was performed using CIAO~4.0.2\footnote{http://cxc.harvard.edu/ciao} \citep{frus06} and the CALDB~3.5.0\footnote{Since data reduction was completed new {\it Chandra} calibrations have been released. The majority of changes relate to calibration updates for recent observations that will not affect the analysis of older observations such as these. Changes to the HRMA effective area in CALDB~4.1.1 will affect a number of routines used for, e.g., spectral fitting in the 0-2~keV range, but due to the low counts of sources studied here and our focus on photometric properties, the differences are small and inconsequential to this study.} calibration files. Data reduction began with the Level~1 event files using the CIAO {\tt acis\_process\_events} tool to perform background cleansing, gain adjustments, and de-streaking of the data. A new Level~2 event file was produced by filtering out events with non-zero status, bad grades (events with grades 1, 5 or 7 were removed), and events with energies outside the range 0.3~-~8.0~keV (above this energy range the high energy particle background rises sharply whilst the instrumental throughput drops). Both observations were searched for background flaring events by studying the integrated count rate across the field, but no such events were found.

To correct for systematic errors in the {\it Chandra} aspect system (typically less than a few tenths of an arcsecond), we performed a preliminary source detection (see Section~\ref{detect} for details of the method) at a high significance to identify bright X-ray sources, aiming to find at least 50 sources per field. We then cross-matched these source lists with the 2MASS catalogue to search for systematic offsets in the positions of sources from the different catalogues. We performed an iterative inspection of positional offsets followed by adjustment of {\it Chandra} source positions and repeated the cross-correlation between catalogues. We continued until the offsets between catalogues were reduced to $< 0\farcs01$. The final offsets were then applied to the aspect solution files for the Level~1 event files using the CIAO tool {\tt reproject\_aspect}, from which new Level~2 event files were then produced. The offsets applied were $\Delta$(RA,~decl.) = (+0\farcs12, -0\farcs18) and (-0\farcs30, +0\farcs18) for the central and north western fields respectively. Based on the offset dispersion for matched sources we estimate the 1$\sigma$ (2$\sigma$) accuracy of the reprojected astrometry to be 0\farcs22 (0\farcs34). The differences between the offsets applied here and those applied by \citet{alba07} can be attributed to updated {\it Chandra} astrometry available at the time of reduction.

\subsection{X-ray source detection}
\label{detect}

\begin{figure*}
\epsscale{1.0}
\plotone{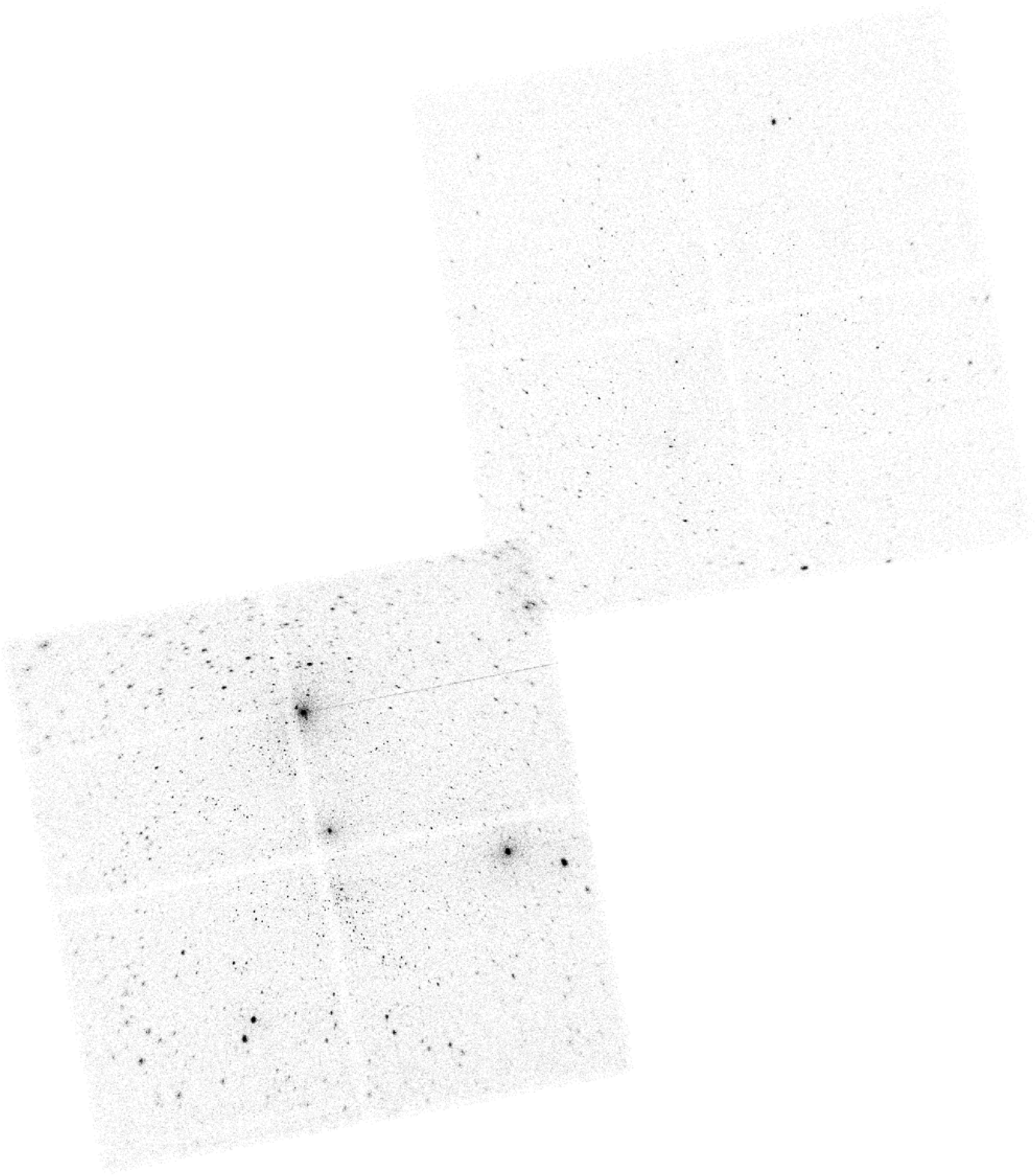}
\caption{Grayscale image of the two {\it Chandra} ACIS-I observations of Cygnus~OB2. The FoV covered by the two observations is approximately 30\arcm~$\times$~37\arcm, with each field 17\arcm~$\times$~17\arcm. The image intensity is proportional to the log of the photon density of the image. The vast number of X-ray point sources present in the Cygnus~OB2 region is evident from this image.}
\label{chandra_image}
\end{figure*}

Figure~\ref{chandra_image} shows an image of the two ACIS-I fields. X-ray source detection was performed using the CIAO {\tt wavdetect} task \citep{free02} on the level two event files. This process works in two stages. First, it detects potential sources by correlating the dataset with a series of ``Mexican Hat" wavelet functions at different spatial scales, identifying pixels with a large correlation value and removing them from the image. The remaining ``cleansed" image is used to estimate the field background, from which detection thresholds are set and tentative sources identified. This process was performed at wavelet scales of 2, 4, 8, 16, and 32 pixels to be sensitive to point-like as well as moderately extended sources or sources at large off-axis angles (this method will miss highly extended sources at large off-axis angles, though it is not the intention of this study to search for such sources). The five resulting source lists were then cross-correlated to remove multiple identifications and a final source list produced.

The separation of reliable and invalid sources is based on the Poisson probability of observing the extracted source counts given the local background density. The {\it significance} of each tentative source, $S_{ij}$, is the probability that the source is not real and is actually a background fluctuation. A {\it significance threshold}, $S_0$, is specified and sources for which $S_{ij} \leq S_0$ are identified as reliable sources. The choice of significance threshold therefore influences the number of spurious noise-based sources which are included (false positives) and the number of real sources which may be missed (false negatives).

We performed source detection at various levels of significance threshold to study the number of sources identified at each. The number of false sources identified can then be estimated as the product of the false source probability density and the size of the field. Using a significance threshold of $10^{-5}$ we detected 990 sources in the central field and 391 sources in the north western field. According to the theory of wavelet source detection, this significance level corresponds to $\sim$40 false sources per field, i.e. $\sim$4\% and $\sim$10\% of detections in the central and north western fields respectively. We labeled these our high-significance sources.

However, wavelet source detection is not as accurate as a full photon extraction and background subtraction in identifying real and false sources and this approach can often fail to detect all the sources in an image. Therefore we repeated the source detection at a significance threshold of $10^{-4}$ and obtained 1993 and 1309 new sources in the two fields. At this significance level a large number of sources detected are likely to be background fluctuations, but some real sources will also be present. Complete photon extraction and cross-correlation with existing source lists at other wavelengths can provide a more stringent separation of true sources and weed out spurious sources, allowing a far deeper source extraction. We labeled these source lists the low-significance sources.
In Section~\ref{s-ccstats} we will show, with the aid of near-IR photometry, that the low-significance sources added to our total source list using this deep source extraction method are predominantly true association members and not background fluctuations. 

Next we performed a visual search of both fields to identify sources missed by both source detection attempts, particularly those in the PSF wings of bright sources. We also added positions for bright optical or near-IR sources known at other wavelengths. This resulted in an extra 48 and 8 sources being added to the low significance source lists of the central and north western fields respectively. Finally, we performed a visual inspection of all sources to search for spurious sources due to CCD gaps, streaks and detector edges and look for multiple detections of the same sources at different spatial scales and at different significance thresholds. 58 sources were removed from the four source lists, which were then merged by field leaving 2993 and 1688 sources in the central and north western fields respectively. Overall this source detection technique varies from that used by \citet{alba07} in that they filtered their source lists based on the detection threshold, while we have performed a deeper source extraction and will filter our source lists based on the Poisson false-source probability, the presence of an association at other wavelengths, or the number of net counts of the X-ray source.

\subsection{Photon energy extraction}
\label{s-photonextract}

Extracting accurate photon numbers and energies from the ACIS event lists is made non-trivial by the non-circular shape and variety of the PSF across the ACIS FoV, particularly at large off-axis angles where it becomes highly asymmetric. A typical PSF can often have highly extended wings, making complete photon extraction impossible in crowded stellar fields.

Photon extraction was performed using ACIS {\sc Extract}\footnote{http://www.astro.psu.edu/xray/docs/TARA/ae\_users\_guide.html} version 2009-01-27\citep[AE,][]{broo02}, an IDL based program for ACIS data processing. Model PSFs were constructed for each source using AE and the CIAO task {\tt mkpsf}, which it correlates with the original source image to refine the source position. 
From the PSF, AE was used to compute polygonal contours for each source that contain a predefined fraction of source events, $f_{PSF}$. This is done at five different energies (0.277, 1.49, 4.51, 6.4, and 8.6~keV) and the values combined to form an aperture correction that is a function of energy. This is then applied to the auxiliary response files (ARFs), correcting the effective area of the observatory as appropriate. 
We used a value of $f_{PSF} = 90$\% for the majority of sources, except those in crowded areas where the value was reduced to prevent overlap with the PSF of other sources. The radius typically increases with off-axis angle due to the deterioration of the PSF.

The standard method for measuring the background for X-ray photon extraction is to construct a circular annulus around each source that avoids the PSFs of other sources and that contains sufficient photons such that the uncertainty in the estimate of net counts is dominated by the source uncertainty and not the background uncertainty. 
This is implemented in AE by  requiring that the background region contain at least 100 background photons and that the background uncertainty for each source is less than the uncertainty in the extracted source counts. 
However, this method has the disadvantage that in crowded regions a local background estimate may be impossible due to the high source density and therefore the measured background can be underestimated. This can seriously bias the extracted properties of weak sources as well as the criteria applied to trim a catalogue built in such a manner. 

AE overcomes these difficulties by implementing a different background measurement technique in crowded regions. The background is treated as the combination of the flat instrumental and sky components and a contribution from the PSF wings of any neighbouring sources. 
The influence of each neighbouring source is assessed in terms of the contribution it has to the number of counts measured in the source and background regions. By iterating over a number of backgrounds models, AE builds up a more accurate estimate of the background than a simple annulus background measurement would provide. 
The requirement that the uncertainty in the final estimate of net counts is dominated by the uncertainty in the extracted source counts is maintained throughout this process, and the ratio between the error associated with the extracted source counts and that of the scaled background subtraction constrained to a minimum value of 4, equating to a maximum photometric error of 3\%. 
Further information on the implementation of this technique in AE (labeled ``better backgrounds'') can be found in \citet{broo02}.

Up until this stage we had extracted sources separately from each {\it Chandra} field, despite the overlap. We then combined the two observations, reprojecting and merging the two event lists and combining the source lists and background estimates. We then performed background-subtracted broad-band X-ray photometry using the combined event lists and extracting from within the PSF appropriate for each observation. Net count rates and fluxes are then estimated with upper and lower limits computed using Gehrel's approximate 84\% confidence limits (equivalent to a 1$\sigma$ Gaussian confidence interval).

The traditional source ``significance'' calculated by {\tt wavdetect} is approximately the photometric signal to noise ratio. Another measurement of the source significance is the Poisson probability, $P_{not}$, that the total counts in the source region could be observed as the chance superposition of background photons. This is effectively the disproof of the null hypothesis that there is no actual source \citep[e.g.,][]{weis07}. By using this method to trim our source catalogue, we not only obtain a suitable measurement of the significance of each source, but also an estimate of the total number of false sources, $N_{false}$,

\begin{equation}
N_{false} \simeq \sum^n_i \, P_{not}
\end{equation}

\noindent and therefore an estimate of the false source fraction, $N_{false} / N_{total}$. We set our criteria for false sources to be $<$~1\% of the total catalogue and for all sources to have a $>$90\% of being real. Therefore we trimmed the catalogue at the $P_{not}$ level to achieve $N_{false} / N_{total} <$~0.01 and $P_{not} > 0.1$. With a significant fraction of sources removed from our source catalogue this way, their photons would now contribute to the overall background levels, lowering the source significance of those sources remaining. Therefore we repeated the photon extraction process with the revised source list and trimmed the resulting catalogue based on the expected fraction of false sources. We iterated over this process until the number of sources remained constant, leaving a total of 1750 sources over the two fields, and an estimated number of false sources, $N_{false} \sim 9$. Of these, 1307 are in the central field, 463 are in the north western field and 20 are in the overlap region. 
We would like to reduce this still further, so in Section~\ref{s-cut} we apply another cut, removing all sources with less than 4 net counts that do not have associations at other wavelengths. By using the presence of associations at other wavelengths we can separate real stellar X-ray sources from background fluctuations and extend the depth of our catalogue. The extracted photon properties of the detected source are given in Table~\ref{table1b}.

\subsection{X-ray spectral analysis}
\label{s-spectra}

To further characterize the properties of the X-ray emitting plasma for these sources we fitted thermal plasma models to the source spectra produced during photon extraction. Detailed spectral analysis of sources in the central Cyg~OB2 field was presented by \citet{alba07} and we reserve discussion of the X-ray spectral properties of sources in the north western field for a further paper. Our objective here is to use results from spectral fitting, such as the hydrogen column density, $N_H$, the plasma temperature, $kT$, and the unabsorbed X-ray flux, $f_X$, to better inform a discussion of the stellar populations in Cyg~OB2. Therefore our goal here is to produce acceptable spectral fits for the majority of sources, but not to fit each source individually or to engage in a discussion of the X-ray emission mechanisms themselves. There is a know degeneracy between plasma temperature and absorption for spectral fits of weak sources \citep[e.g.,][]{flac06}, which may result in overestimated values of $N_H$ and underestimated plasma temperatures. We therefore only perform model fitting for sources with at least 20 net counts.

Spectral fitting was performed using {\sc xspec}\footnote{http://heasarc.nasa.gov/docs/xanadu/xspec/} version 12.4.0 \citep{arna96} and compared to {\sc apec} \citep{smit01} model spectra corresponding to single-temperature thermal plasma in collisional ionization equilibrium. The elemental abundances were frozen at 0.3~Z$_{\odot}$ \citep[a level which has been suggested from various studies of star-forming regions; e.g., ][]{tsub98,hama00,tsub00}. Interstellar X-ray absorption, characterized by the hydrogen column density, $N_H$, was modeled using {\sc tbabs}\footnote{http://astro.uni-tuebingen.de/nh/} \citep{balu92}. The photoelectric absorption cross-sections and solar abundance tables used for {\sc tbabs} are described in \citet{wilm00}. 

In fitting models to our spectra we employed the C-statistic to determine goodness-of-fit, a method more appropriate to low-count data than using the $\chi^2$ statistic \citep{cash79}. The C-statistic is an application of the likelihood ratio test and involves computing the likelihood of the combined source and background X-ray events being observed. This has the advantage that the background does not need to be subtracted from the observed counts, but it must be modeled and combined with the source model to be fit. We use the well-constrained {\it cplinear} background model included within ACIS {\sc Extract} as opposed to the traditional \citet{wach79} background model which is poorly constrained for sources with low counts \citep{broo02}. The combined spectrum then passes through the {\it Chandra} ACIS instrumental response and is compared to the source spectrum via the C-statistic. Because application of the C-statistic does not require the data to be rebinned (as with the $\chi^2$ method), it is particularly appropriate for sources with small numbers of counts. 

We set up a grid of initial parameter values covering $N_H$~=~1~--~5~$\times 10^{22}$~cm$^{-3}$ and $kT$~=~0.7~--~2.0. For each source we attempted a fit starting with the initial parameters for each point on the grid. We then chose the best fitting spectral model from all the fits. This allows us to avoid accidentally fitting relative minima in C-statistic space. {\it Chandra} has a reduced sensitivity to X-rays above 8-10~keV due to the rapid decline in telescope effective area. Therefore, for all model fits with plasma temperatures above 10~keV we truncate the value to 10~keV, and these are listed in the tables as ``$> 10$~keV''.

Fits were found for all 635 X-ray sources with more than 20 net counts. We chose not to perform more advanced spectral fitting, such as using two-temperature models, since our focus is on stellar properties for studying the overall Cyg~OB2 population and not the X-ray spectral properties of the most massive stars \citep[][presented an analysis of the X-ray model fits for OB stars in the center of Cyg~OB2]{alba07}. For these sources we find that the distributions of hydrogen column densities peak at log~$N_H = 22.16$ and 21.98~cm$^{-2}$, for the central and north western fields respectively. Using the conversion $N_H = 2.2 \times 10^{21} A_V$~cm$^{-2}$ \citep{ryte96} gives $A_V = 6.6$ and 4.3 for the two fields. The extinction in the central field is in good agreement with that estimated by previous studies in the optical and near-IR \citep[e.g.,][]{mass91,knod00}, while in the north west a well known reddening hole in this area \citep{redd67,drew08} allows a greater line of sight with the extinction only reaching levels of $A_V \sim 4-6$, in agreement with our results. The modal value for the plasma temperature in both fields, $kT \sim 2$, is typical of young stellar sources and in agreement with studies of other SFRs \citep[e.g.,][]{getm02}.

For sources with less than 20 net counts we use a number of other methods to derive useful X-ray properties. 
To estimate their X-ray fluxes we calculated a conversion factor between count rate and $f_X$ based on the mean ratio between unabsorbed X-ray flux and source count rates from all the sources with model fits. This conversion was found to be 2.15~$\times 10^{-11}$~erg. 
We then applied this conversion to the fainter sources to derive X-ray fluxes. The errors on these X-ray fluxes were derived from the errors on the net count rates and the dispersion of the count rate to $f_X$ conversion used.

To determine $N_H$ for these faint sources we employ the fact that the background-subtracted median energy of a source's photons is a reliable indicator for the absorbing hydrogen column density \citep[e.g.,][]{hong04,feig05}. 
\citet{feig05} derived a relationship between median energy and $N_H$ from {\it Chandra} Orion Ultradeep Project observations, allowing $N_H$ to be determined for sources with only a few counts. We have used this relationship to derive hydrogen column densities for all our faint sources based on their median energies. 
The dispersion of this relationship at low median energies means that for sources with median energies $< 1.3$~keV it is not possible to derive an accurate value of $N_H$ and therefore such sources have been labeled with ``$< 22$''.

The results of X-ray spectral fitting and derivation of properties for fainter sources are listed in Table~\ref{table2b} for all sources in our catalogue (excluding those removed following the cut described in Section~\ref{s-cut}).

\subsection{Variability analysis}
\label{s-variable}

In this section we attempt to quantify the variability of the observed X-ray emission with particular attention to determine quantities which may aid studies of the stellar populations. Young, low-mass X-ray sources are known to show high levels of variability with various origins including magnetic flares and rotationally modulated emission \citep[e.g.,][]{flac05}. All coronally emitting X-ray sources are believed to exhibit flare-like variability given a long enough observation \citep{getm05,cara07} and observations of such behavior can confirm a stellar origin for the X-ray source. High-mass sources are less variable, the suggested origin of their X-ray emission from multiple small shocks in their strong stellar winds not giving rise to such large fluctuations. Extragalactic X-ray sources may also show variability, though with different characteristics.

Light curves were produced for each source and binned into bins of unequal width, but constant significance. X-ray variability was then investigated using a one-sample Kolmogorov-Smirnov test to compare the distribution of photon arrival times with that expected for a constant source (the null hypothesis). Sources with a probability of accepting the null hypothesis of log($P_{KS}) > -1.3$ ($>$~5\%) were classified as ``constant'' and sources with log($P_{KS}) < -2.3$ ($<$~0.5\%) were classified as ``variable''. Sources between these two limits were classified as ``possibly variable''.

This method identified 223 sources (13\%) as variable and 152 sources (9\%) as possibly variable (though approximately half of the ``possibly variable'' sources are expected to be false positives). These fractions are likely to be lower limits due to both the sensitivity of the KS test on photon statistics (i.e. short events such as flaring are harder to identify in sources with small numbers of counts due to the increasing time bin width for weak sources) and the chances of not observing a flaring event during a finite observation. Despite this, classification of a source as variable can be useful in identifying cluster members if the form of the variability can be identified as a stellar flare event. The results of X-ray variability analysis for all sources in our catalogue (excluding those removed following the cut described in Section~\ref{s-cut}) are listed in Table~\ref{table1b}.

\subsection{Catalogue sensitivity limits}
\label{s-limits}

\citet{feig05} derived an estimate for the on-axis {\it Chandra} point source sensitivity from the Orion Ultradeep Project as

\begin{eqnarray}
\mathrm{log} \, L_x = 28.7 \, + & \, 2 \, \mathrm{log} \, (d / \mathrm{kpc}) \, - \, \mathrm{log} \, (t_{exp} / 100 \mathrm{ks}) \, \nonumber \\
& + \, 0.4 \, (\mathrm{log} \, N_H \, - \, 20) \, \mathrm{erg~s}^{-1}.
\end{eqnarray}

\noindent Assuming a distance of 1.45~kpc \citep{hans03} and using the mean hydrogen column densities determined above, we derive sensitivity limits of log$\, L_x = 29.9$~erg~s$^{-1}$ and 30.1~erg~s$^{-1}$ in the central and north western fields respectively. This implies an at-telescope sensitivity of 3.1~$\times 10^{-15}$~erg~cm$^{-2}$~s$^{-1}$. 

The faintest on-axis sources emerging from the source detection and extraction procedure have net count rates of $\sim$2$\times 10^{-5}$~s$^{-1}$. The corresponding minimum detectable X-ray flux in the 0.5-8~keV band is $f_X \simeq 6.9 \times 10^{-16}$~erg~cm$^{-2}$~s$^{-1}$ (calculated using PIMMS\footnote{The {\it Chandra} Portable Interactive Multi-Mission Simulator, http://cxc.harvard.edu/toolkit/pimms.jsp} and accounting for the extinction, exposure time, and a typical source spectrum with $kT = 2$). The deep source extraction routine used here has therefore allowed us to detect sources with fluxes lower than the sensitivity limit estimated by \citet{feig05}.

\section{Optical and near-IR cross-correlation}
\label{s-cc}

In this section we cross-correlate the X-ray source list with point source catalogues (PSCs) from recent deep photometric optical and near-IR (OIR) surveys covering the Cygnus~OB2 region. Association members are likely to have optical or near-IR counterparts, while false sources and extragalactic background contaminants are less likely to have associations at other wavelengths. 
Therefore, while the identification of an OIR association for and X-ray source is not proof that it is a stellar X-ray member of Cygnus~OB2, it does lend strong credence to the source being real. 
This cross-correlation will therefore allow us to weed out spurious detections and increase our true source fraction as well as provide vital photometric information on these sources.

\subsection{The optical and near-IR catalogues}
\label{s-cats}

We employ data from three deep large-area surveys with coverage in the Cygnus region: the INT Photometric H$\alpha$ Survey \citep[IPHAS,][]{drew05}, 2MASS \citep{cutr03}, and the UKIRT Infrared Deep Sky Survey \citep[UKIDSS,][]{lawr07} Galactic plane Survey \citep[GPS,][]{luca08}. IPHAS is a 1800 deg$^2$ photometric survey of the northern Galactic plane (latitude range $-5^{\circ} < b < +5^{\circ}$) in Sloan $r'$, $i'$, and narrow band H$\alpha$ filters. IPHAS data are capable of separating differently reddened line-of-sight stellar populations \citep[e.g.,][]{drew08,sale09} and is therefore especially useful for studying SFRs in the Galactic plane. The survey has particularly deep dark-time data for the Cygnus region, complete to $r' = 21$.

Source catalogues for the Cygnus~OB2 region were obtained through the AstroGrid VO Desktop\footnote{http://www.astrogrid.org/} from the IPHAS Initial Data Release \citep{gonz08}. We find 2800 and 5789 IPHAS stellar sources in the central and north western fields, respectively. The vast difference in source counts between the two regions is a function of the considerably different extinction down the two sight lines (see Section~\ref{s-spectra}). Assuming a distance of 1.45~kpc and a 2~Myr-old PMS population, we find that IPHAS should be complete down to masses of $\sim$1.2 and $\sim$0.65~M$_{\odot}$ for the central and north western fields respectively.

\begin{figure}
\epsscale{1.2}
\includegraphics[width=280pt, angle=270]{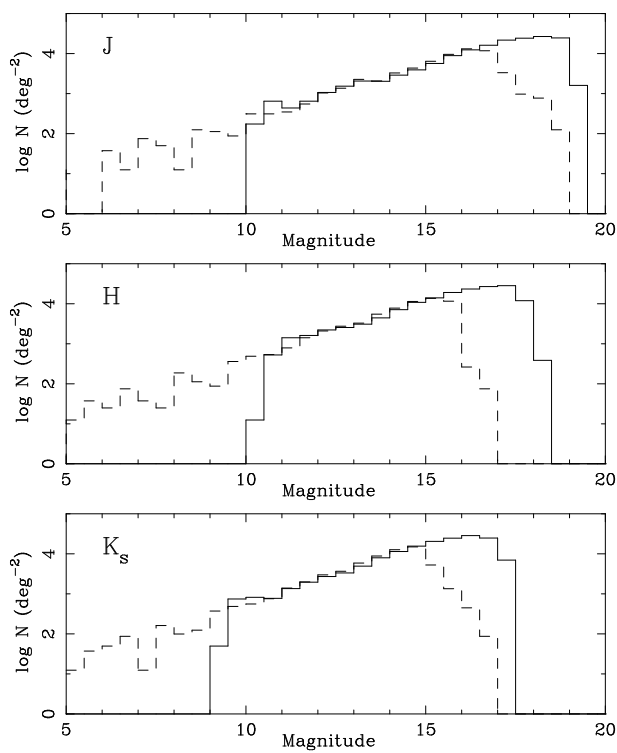}
\caption{$J$, $H$, and $K_s$ band star counts from the 2MASS (dashed line) and UKIDSS (full line) near-IR catalogues for the area of the {\it Chandra} observations. The UKIDSS photometry was converted onto the 2MASS photometric system for comparison. To produce a unified near-IR catalogue the two source lists were cross-correlated and magnitudes taken from 2MASS data in the bright limit and UKIDSS data in the faint limit. In the overlapping magnitude range we chose the observations with the lowest photometric error.}
\label{nearIRmags}
\end{figure}

In the near-IR regime we use data from 2MASS and UKIDSS, both of which combine photometry from the $J$, $H$, and $K$ bands using slightly different filter combinations \citep[][provide a list of transformations for colors and magnitudes between the two systems]{hewe06}. 2MASS is complete to a depth of $K_s \sim 14.5$ in the Cygnus region, while UKIDSS-GPS extends two magnitudes deeper to $K_s \sim 16.5$ (Figure~\ref{nearIRmags}). Because UKIDSS observations saturate below $K_s \sim 10$, we require observations from both catalogues for complete coverage.

We find 5126 and 5265 stellar 2MASS sources in the central and north western fields respectively. We select only sources with valid detections (photometric quality flags of A, B, C or D) in one of the three bands. In his analysis of 2MASS observations of Cygnus~OB2, \citet{knod00} estimated that the photometry for a main-sequence population at 1.7~kpc was complete down to a mass of $\sim$1.5~M$_{\odot}$ \citep[for a 2~Myr PMS population with DM=10.8 and $A_V = 7$ this would equate to a depth of $\sim$0.95~M$_{\odot}$;][]{sies00}.

At the time of writing the UKIDSS-GPS is not complete and observations in the Cygnus region are only partly finished, with a small part of the north western field not yet observed. Despite this we find a source density almost three times that of 2MASS with 13,184 stellar sources identified in the central field (with photometric errors $< 0.1$ in all three bands). This high source density is testament to the incredible depth of the UKIDSS-GPS, which will allow PMS stars down to $\sim$0.20~M$_{\odot}$ to be detected in the Cygnus~OB2 region. We anticipate that the lack of complete coverage in the north western field will not be a serious drawback due to the lower X-ray depth of the observations there.

For photometric simplicity we converted all UKIDSS measurements onto the 2MASS system using the transformations given in \citet{hewe06}. Figure~\ref{nearIRmags} shows star counts for the two surveys for the areas of the {\it Chandra} observations and there is a good agreement in source counts in the overlapping magnitude range. To produce a unified catalogue we cross-correlated the two PSCs to produce a single near-IR source catalogue, using 2MASS magnitudes in the bright limit, UKIDSS magnitudes in the faint limit and in the overlapping magnitude range we chose the observation with the lowest error (often UKIDSS). We further cross-correlated this catalogue with the IPHAS data for the region to produce a complete OIR catalogue for the region of the two {\it Chandra} fields. This process was made simple due to the fact that all three surveys share the same astrometric system and have very high positional accuracies (global astrometric precisions are 0\farcs15 for 2MASS, 0\farcs25 for IPHAS, 0\farcs3 for UKIDSS), and therefore cross-correlation radii of only 0\farcs5 were necessary, reducing any risk of miss-associating sources. In total there are 8327 optical sources, 20,364 near-IR sources, and 22,150 unique OIR sources in the two fields.

\subsection{Cross-correlation method}
\label{s-ccmethod}

The high spatial density of OIR sources in the Cygnus~OB2 cluster combined with the variable and non-gaussian PSF of {\it Chandra} sources at high off-axis angles requires a flexible but self-consistent method of identifying OIR counterparts for each X-ray source that depends on the quality of the X-ray astrometry and PSF. A compromise must be made between maximizing the number of associations found and limiting the contamination from unrelated sources. The use of extremely populous source catalogues such as UKIDSS may not only produce a high fraction of true associations but also increase the risk of making a false match.

To accurately identify the correct OIR counterpart for each X-ray source we adopt a Bayesian scheme similar to that outlined in \citet{bran06}. The method uses information on the magnitude distribution of OIR sources to estimate the probability of an OIR source of a given magnitude lying at a given distance from an X-ray source by chance, and therefore its probability of being the X-ray source's true OIR counterpart. The method also calculates the probability that the source has no identifiable OIR association down to the limiting magnitude of the survey. This Bayesian scheme is more advanced than a simple matching radius method because it can self-consistently include additional information contained within the datasets themselves, such as the source densities, positional errors, and dispersion of the OIR catalogues. Our approach differs from that implemented in \citet{bran06} because we have both optical and near-IR catalogues, each with their own source density and properties and we have therefore performed the Bayesian search method for each catalogue and combined the results to assess the probability of each source in our merged OIR catalogue.

The Bayesian source identification method is fully described in \citet{bran06}, but we repeat the main equations here. The probability of identifying X-ray source $i$ with OIR source $k$ is given by

\begin{equation}
P_{ik} = f \, \frac{M_{ik}}{B_k} \left[ (1-f) \, + \, f \, \sum_{l=1}^{n_i} \frac{M_{il}}{B_l} \right]^{-1}
\end{equation}

\noindent and the probability of source $i$ having no association in that OIR catalogue is

\begin{equation}
P_{i0} = (1-f) \, \left[ (1-f) \, + \, f \, \sum_{l=1}^{n_i} \frac{M_{il}}{B_l} \right]^{-1}
\end{equation}

\noindent where $f$ is the estimated mean fraction of sources with OIR associations and is a Bayesian parameter which is marginalized over. $M_{il}$ is a Gaussian model of the probability of two sources being associated with each other, and incorporates the separation of the two sources, the X-ray positional error (based on the degradation of the {\it Chandra} PSF from the optical axis), and the dispersion of the OIR catalogue. $B_l$ is the differential number counts of sources, $B = dN / dm$, at the magnitude of source $l$. The sum is performed over all sources and factors in the probability of finding a source of magnitude $l$ in that area. Identification probabilities are then estimated by marginalizing over $f$ and normalizing.

Once all the identification probabilities are computed, as well as the probability of the source not having an identifiable association in the OIR catalogues, the source with the highest probability is identified as the counterpart to the X-ray source. We also impose a maximum positional offset that if the identified source exceeds it is discarded and the X-ray source left without a counterpart. As the maximum separation radius for associating an OIR source with an X-ray source, we use the combined positional uncertainties of the OIR catalogues and the 95\% confidence level {\it Chandra} positional uncertainty derived by \citet{kim07} from the {\it Chandra} Multi-wavelength Project PSC

\begin{equation}
\mathrm{log} \, \delta_X = \left\{ 
\begin{array}{ll}
0.1145 \, \theta - 0.4958 \, \mathrm{log} \, C_N &+ 0.1932\\
&\mathrm{for} \, C_N < 137.82 \\
0.0968 \, \theta - 0.2064 \, \mathrm{log} \, C_N &- 0.4260\\
&\mathrm{for} \, C_N > 137.82\\
\end{array} \right.
\end{equation}

\noindent where the positional uncertainty, $\delta_X$, is in arcsec, the off-axis angle, $\theta$, is in arcmin, and $C_N$ is the number of net counts for the X-ray source.

\begin{figure}
\includegraphics[width=130pt, angle=270]{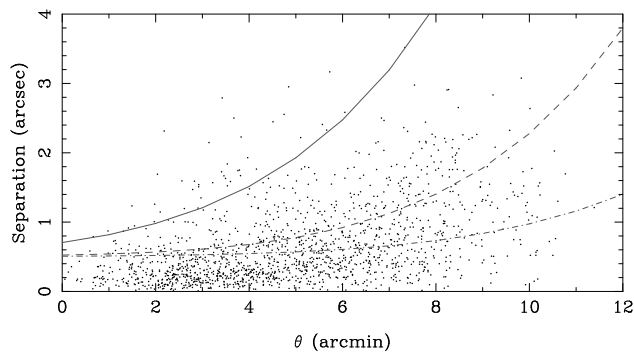}
\caption{Separation between cross-matched X-ray and OIR sources for all 1501 cross-correlated {\it Chandra} sources shown against the off-axis angle of the X-ray source. Also shown are the maximum separation radii as a function of off-axis angle for sources with 10 (full line), 100 (dashed line) and 1000 counts (dot-dashed line). The mean separation distance is 0\farcs65, with the majority of sources (79\%) within 1\arcs.}
\label{offsets}
\end{figure}

Using these methods we find likely OIR counterparts for 1501 of the 1750 X-ray sources in our source list (86\%), the majority of these (79\%) within 1\arcs\ of the X-ray source. Figure~\ref{offsets} shows the separation between associated X-ray and OIR sources as a function of the {\it Chandra} off-axis angle. Of these matches, 1490 are the closest OIR source to the X-ray source, while 11 are the second closest source, but which are less likely to be there by chance, so are chosen by the Bayesian method as the more likely counterpart. 
An example of this is X-ray source number 418. The closest OIR source to it is a UKIDSS source with a magnitude of $K_s = 14.9$. Given the high number of sources this faint in the field it has a relatively high probability of being there by chance. Slightly further away is another UKIDSS source with a magnitude of $K_s = 11.2$. Sources of such a high brightness are much rarer and are less likely to be there by chance. This combination means that it is statistically more likely that the X-ray source is associated with the brighter source and that the fainter source is there by chance, than vice versa. The Bayesian technique therefore associates the X-ray source with the brighter OIR source. 
Of the 249 unmatched X-ray sources, 29 have OIR counterparts within the maximum separation radius, but for which the probability of the source being the true counterpart is lower than the X-ray source having no identifiable association. We do not find any significant difference in the fraction of X-ray sources with associations for low and high significance X-ray source detections, which supports our assumption that the majority of these sources are real. 
Results of the X-ray~--~OIR cross-correlation for all sources in our catalogue (excluding those removed following the cut described in Section~\ref{s-cut}) are listed in Table~\ref{table3b}. 

\subsection{A Monte Carlo simulation of the cross matching technique}
\label{s-mc}

To quantify the reliability of our cross-matching technique and determine the likely fraction of false associations we ran Monte Carlo simulations of the OIR and X-ray observations to model this problem. The existing X-ray catalogue of positions and net counts was used and we assumed that a fraction, $f_{mc}$, of the X-ray sources had true counterparts in one of the OIR catalogues. The position of this source was randomly drawn based on a Gaussian model for the relative positional uncertainty \citep[e.g.,][]{bran06} characterized by

\begin{equation}
\sigma^2 = \sigma_{OIR}^2 + \frac{1}{C_N} \left[ \sigma_0 + \sigma_{600} \left( \frac{d_k}{600\arcs} \right)^2 \right]
\end{equation}

\noindent where $\sigma_{OIR}$ is the uncertainty in the OIR position, $\sigma_0 = 0\farcs6$ is the {\it Chandra} PSF width at the pointing center ($d_k = 0$), and $\sigma_{600}$ describes the quadratic growth of the PSF as a function of the off-axis angle, $d_k$). $C_N$ is the number of net counts for the X-ray source. The photometric properties of the source were randomly chosen from the stellar catalogue of the region. Background OIR sources were then distributed randomly across the field using the known spatial source density until the number of OIR sources matched those in the observed catalogues. The Bayesian matching technique was then applied to the modeled source catalogues exactly as described above.

The results of this simulation tell us how many X-ray sources will be associated with an OIR source, given the value of $f_{mc}$ used. We therefore iterated over $f_{mc}$ until a value was found that reproduced our result of finding associations for 86\% of sources. Since the cross matching technique is dependent only on the properties of the OIR sources and the dispersion and positional accuracies of the catalogues, but not on the properties of the X-ray sources, we may use this to determine the underlying fraction of sources with true OIR counterparts based on the number of counterparts we find using the Bayesian technique. This not only tells us what the true fraction of sources with associations actually is but how many ``false positives'' and ``false negatives'' we should expect within our sample. The results of this simulation are given in Table~\ref{mcresults} as ``Simulation 1''. 
In summary, a true fraction of $f_{mc} = 0.85$ of sources having real OIR associations will result in the Bayesian matching scheme finding associations for 86\% of sources. The two largest sources of error are ``false-positives'' ($\sim$13\% of unassociated X-ray sources will be falsely associated with an OIR counterpart) and mis-associations ($\sim$6\% of X-ray sources with OIR associations will be incorrectly associated with a different OIR counterpart\footnote{This fraction is a product of the 13\% chance of any X-ray source having an OIR source within its separation radius and the fact that X-ray sources with true associations will also have their real OIR counterpart within their separation radius as well. The Bayesian matching scheme appears to be able to identify the correct OIR source 56\% of the time. Since this is only based on the magnitude of the OIR source, we consider this to be a good result.}). Applying these statistics to our 1501 real X-ray~--~OIR matches we estimates that $\sim$12 are ``false-positives'' and $\sim$92 have been miss-associated. In total, 93\% of our 1501 cross-matches are expected to be correct.

The result that can be drawn from this is that the greatest source of error in matching X-ray sources with OIR sources from such deep and populous catalogues as IPHAS and UKIDSS is due to identifying the wrong OIR source with the X-ray source. This is clearly a product of the positional uncertainty of {\it Chandra} sources, particularly at large off-axis angles, when compared to the recent generation of deep and high spatial resolution photometric catalogues in the optical and near-IR.

\subsection{A catalogue cut based on source association}
\label{s-cut}

Our catalogue contains many sources with a low number of counts, which may include a number of false sources. By summing $P_{not}$ over the entire catalogue we have estimated that amongst our 1750 X-ray sources there are $\sim$9 false sources. Identifying the false sources and removing them is often not possible based on the X-ray observations alone and therefore observers often apply a conservative cut level to reduce the false source fraction. 
Another technique for separating false sources from real sources is to use information at other wavelengths. Young stellar X-ray sources at the distance of Cygnus~OB2 should have detectable associations in the near-IR, especially considering the depth of the catalogues we have used (e.g., UKIDSS observations are complete to 0.2~M$_{\odot}$ in this region, see Section~\ref{s-cats}). 
Therefore false sources such as background fluctuations will be confined to those sources without identified OIR associations (excluding $\sim$13\% of sources which may have chance OIR associations, see Section \ref{s-mc}), while true stellar X-ray sources will have identifiable OIR associations (excluding a fraction $\sim$1\% for which the OIR counterpart could not be identified).

To implement this as a cut to our catalogue and hopefully remove a large fraction of the remaining false sources we removed all X-ray sources with less than 4 net counts and without OIR associations. 
Of the 186 sources in our catalogue with less than 4 net counts, 132 (71\%) have OIR associations and 54 do not. 
To find out what this tells us about the number of these sources with true OIR counterparts we repeat the Monte Carlo simulation described in Section~\ref{s-mc}, iterating until the simulation matches the same fraction of sources as we have found. The results of this are listed in Table~\ref{mcresults} as ``Simulation 2'' and tell us that 69\% of the 186 sources with less than 4 counts have real OIR associations, and that we have correctly identified their counterpart in $\sim$91\% of cases. Of the 54 sources we have discarded, $\sim$88\% are truly unassociated sources and the rest are ``false negatives''.

This shows that by filtering weak X-ray sources based on the presence of a counterpart at another wavelength we may maintain the majority of true X-ray sources and dispose of many false sources. The final catalogue contains 1696 sources with 1501 (89\%) having associations in the optical or near-IR. Of these 1277 are in the central field, 439 are in the north western field and 20 are in the overlap region. 
A small fraction of the retained sources (and of the brighter X-ray sources) may be background extragalactic sources for which we have associated the source with its actual OIR component. While this contamination fraction is harder to quantify we will show in Section~\ref{s-ccstats} that the near-IR properties of both low- and high-significance X-ray sources are similar and in agreement with that expected for the young stellar population in Cyg~OB2.

\subsection{Correlation with other catalogues}
\label{s-morecats}

Due to the large number of OB stars in Cyg~OB2 there have been many spectral surveys of the region designed to uncover the majority of these massive stars. We cross-correlated our X-ray catalogue with these spectroscopic catalogues in the literature, finding sources within our FoV from \citet{mass91}, \citet{herr99}, \citet{come02}, \citet{kimi07}, and \citet{negu08}. 
We identified 61, 1, 5, 53, and 1 source, respectively, from each of these catalogues, and found X-ray associations for 37, 1, 0, 15, and 1 source. 
These matches include all 29 O-type stars in the FoV, but only 25 of the 87 known B-type stars in the FoV. This division in observed X-ray emission between O-type (and early B-type) stars and late B-type stars is expected based on the origin of X-ray emission in massive O-type stars from strong stellar winds and the lack of such winds in late B-type stars. X-ray emission observed from B-type stars is often attributed to the presence of a lower-mass binary companion

From the near-IR spectra presented in \citet{come02} we do not associate X-ray sources with either of the two objects in our FoV suggested to be O-type stars based on their featureless near-IR spectra. The O7.5{\sc i} star suggested to be an O-type by \citet{come02} and identified as such by \citet{negu08} is detected however. 
We also do not find X-ray associations for three other sources identified by \citet{come02} with either Br~$\gamma$ or CO emission and suggested to be evolved massive stars.

We have also matched our detected sources with those identified by \citet{alba07} in the central {\it Chandra} field, finding 992 matches between our catalogue and the 1003 sources they identified. The 11 sources detected by \citet{alba07} and not detected by us are all very faint and this difference between the catalogues is a product of the different detection and reduction techniques employed and the different criteria used to trim the catalogues. We attempted to include these missing sources by adding their positions to the source list we compiled using {\tt wavdetect}, but their not-a-source probabilities were all above the level we have used as a cutoff for our catalogue, so they were not included.

The results of cross-matching our sources with those from other catalogues are listed in Table~\ref{table3b}.

\subsection{Sources without optical or near-IR associations}

195 of the 1696 X-ray sources in our catalogue do not have OIR associations, the majority with less than 20 net counts. These are likely to be a combination of highly obscured PMS association members, extragalactic sources such as quasars and other active galactic nuclei (AGN), and background fluctuations. There are four bright unassociated X-ray sources with more than 100 net counts that are worthy of individual attention. Source no.~97 with $\sim$137 net counts has no identifiable counterpart in IPHAS or 2MASS images, but the lack of UKIDSS coverage in its vicinity may be responsible for the lack of identification. Source no.~134 with $\sim$568 net counts is one of the brightest X-ray sources in the observed region. With such a high number of counts and an off-axis angle of $\sim$7$^{\circ}$ the source has a positional uncertainty of $< 1$\arcs. The closest OIR source is 5\farcs2 away, likely too far to be associated with this source. 
X-ray sources 1003 and 1196 have $\sim$247 and $\sim$383 net counts respectively. Inspection of deep UKIDSS imagery reveals they are both associated with faint near-IR sources blended with brighter sources that have their own X-ray association. Since their photometry cannot be extracted we are forced to leave them unassociated, though we believe them to be stellar in origin.

These two unassociated sources could be highly embedded massive stars, or background galactic or extragalactic sources. To distinguish between the two we model both using a $\Gamma = 1.4$ power law to derive their X-ray fluxes. Our OIR observations for these two sources are limited to the 2MASS completeness limit of $K_s \sim 15$. At this magnitude limit both sources lie in the region typically occupied by AGN in the $K$ versus $F_X$ diagram from \citet{brus05}. UKIDSS observations of this region will be necessary to confirm or deny their identity as AGN. If they are not detected by UKIDSS they are likely to be optically too faint to be AGN and would then more likely to be heavily embedded stellar sources.

The potential number of extragalactic sources contributing to our X-ray source list may be determined by estimating the sensitivity of our observations to extragalactic sources. For the extragalactic sources we assume a power-law spectrum with a photon index, $\Gamma = 1.4$, and a hydrogen column density derived from the maximum Galactic extinction in each field\footnote{We were forced to derive the total hydrogen column density from the maximum Galactic extinction, $A_V$, because $N_H$ values derived directly from H~{\sc i} surveys using the Colden calculator (part of the {\it Chandra} proposal planning toolkit) were smaller than the mean hydrogen column densities for our sources, and therefore likely to be inaccurate.} \citep{schl98} and the conversion $N_H = 2.2 \times 10^{21} A_V$~cm$^{-2}$ \citep{ryte96}. Using PIMMS and the mean net photon count of unassociated sources in each field we derive limiting extragalactic sensitivities of $2.0 \times 10^{-15}$ and $3.0 \times 10^{-15}$~erg~cm$^{-2}$~s$^{-1}$ (in the hard band, 2-8~keV) for the central and north western fields respectively (the lower sensitivity to extragalactic sources compared to the stellar sensitivity determined in Section~\ref{s-limits} is due to the high total Galactic extinction behind Cyg~OB2, which \citealt{schl98} estimate as $A_V \sim 18$).

Using the extragalactic source density as a function of sensitivity \citep[derived from {\it Chandra} Deep Field North observations,][]{bran01} we estimate the number of extragalactic sources contributing to our source lists as $\sim$113 and $\sim$75, for the central and north western fields respectively. We estimated in Section~\ref{s-mc} that $\sim$13\% of truly unassociated source are likely to have false associations identified for them using our Bayesian scheme, indicating that $\sim$24 extragalactic X-ray sources are likely to have been included within our source list. The remaining $\sim$164 unassociated extragalactic sources are likely to dominate the unassociated X-ray sources.

Distinguishing between embedded stellar sources and extragalactic sources is not simple. Spectral fits using stellar thermal plasma or extragalactic power-law models might reveal the more likely origin, but the fact that the majority of these sources are faint ($< 20$ net counts) means that model fits would be unreliable. Stellar sources may also be identified if their X-ray light curves exhibit distinctive flaring events indicative of coronal activity \citep[e.g.,][]{gros04,flac05}. We identify 17 unassociated X-ray sources which were classified as `definitely variable' from their light curve (Section~\ref{s-variable}), of which 9 have more than 20 net counts. Five of these show variability in the form of a rapid high-amplitude rise and a slower decay. We suggest that these are likely stellar sources, either embedded or beyond the limits of the OIR catalogues (later than type M6), with notably strong X-ray emission. The remaining unassociated sources are therefore likely to be a combination of extragalactic sources, background fluctuations or stellar sources that did not exhibit flaring during the observations.

\section{A statistical analysis of the deep source extraction routine}
\label{s-ccstats}

\begin{figure*}
\includegraphics[width=370pt, angle=270]{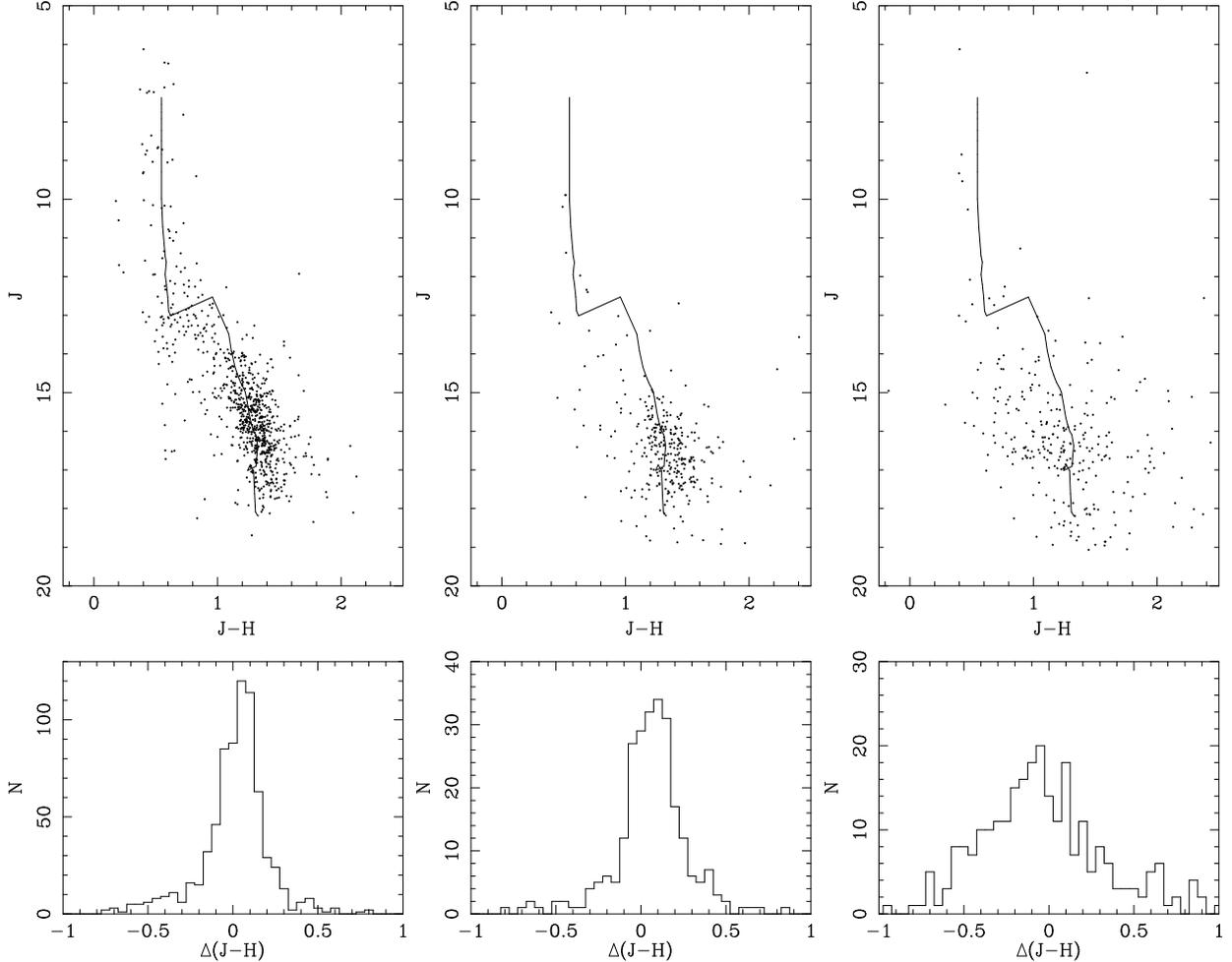}
\caption{{\it Top:} near-IR color-magnitude diagrams for three samples of X-ray sources from the central {\it Chandra} Cyg~OB2 field. 
{\it Left:} the 814 high-significance sources. 
{\it Center:} the 304 low-significance sources. 
{\it Right:} a Monte-Carlo simulation of 304 X-ray sources in the central {\it Chandra} field. 
Shown for reference are $A_V = 7.0$, 2~Myr pre-MS isochrones \citep[from][]{sies00} at the distance of Cygnus~OB2 (DM=10.8). {\it Bottom:} histograms of the distribution of $(J-H)$ around the 2~Myr pre-MS isochrones for each sample (excluding the degenerate region around $12.5 < J < 13$). A Kolmogorov-Smirnov test of the $\Delta (J-H)$ color distribution indicates an 89\% probability that the low-significance sources are drawn from the same sample as the high-significance sources and a $< 1$\% probability that the low-significance sources are drawn from the random source list. This, and the similar and higher fraction of sources with OIR associations of the high- and low-significance source lists compared to the randomly-positioned X-ray sources supports our assertion that the deep source extraction routine that produced the low-significance source lists has produced a valid source list of stellar X-ray sources.}
\label{montecarlo_cmds}
\end{figure*}

In this section we intend to show that the high and low-significance sources detected in Section~\ref{detect} are drawn from the same population and that the low-significance sources with OIR associations are not background fluctuations, as might be expected for many faint X-ray sources. To do this we will compare the near-IR properties of the high and low-significance sources with those drawn randomly from the field population in the direction of Cyg~OB2. For the purpose of this we consider only sources in the central {\it Chandra} field, due to the different extinction in the two fields. In Figure~\ref{montecarlo_cmds} we show near-IR color-magnitude diagrams for the high-significance (814 sources), low-significance (304 sources) and a Monte Carlo near-IR source list (304 sources). Also shown are reddened (at $A_V = 7$) 2~Myr pre-main sequence (PMS) isochrones from \citet{sies00}. Both the low and high significance X-ray sources are distributed around the PMS isochrone, with the low significance sources clustered more toward the fainter end of the isochrone (as would be expected since low-mass stars are generally less X-ray luminous than coeval high-mass stars). The Monte Carlo sources show a much more random distribution representative of the color-magnitude distribution for all sources in this sight-line, including foreground and background sources.

The ($J-H$) color distributions around the 2~Myr PMS isochrone can be seen in the lower panels of Figure~\ref{montecarlo_cmds} (excluding the degenerate region around $12.5 < J < 13$ where such a calculation isn't possible). The distribution of low-significance sources closely resembles that of the high-significance sources, while the random sources show a much wider spread. To quantify this association we performed a two-sample Kolmogorov-Smirnov (K-S) test on the $\Delta(J-H)$ color distribution. We find an 89\% probability that the low-significance sources are drawn from the same sample as the high-significance sources and $< 1$\% probability that the low significance sources are drawn from the random source list. We assert that this, and the high cross-correlation fraction for these sources compared to randomly positioned X-ray sources is evidence that the low significance sources detected here are valid stellar X-ray members of Cyg~OB2.

\section{Summary}

We have presented a catalogue of 1696 X-ray sources which were extracted from two archival {\it Chandra} observations of the center of the massive SFR Cygnus~OB2. Sources were detected using the wavelet-based CIAO {\tt wavdetect} tool and analyzed using the routines in ACIS {\sc Extract}. This includes extraction of broad band X-ray fluxes, spectra, and light curves. The spectra were fitted using single temperature thermal plasma models and the light curves assessed for variability using a Kolmogorov-Smirnov test.

We adopted a deep source detection routine to maximize the number of sources extracted. To limit the number of false sources identified as real sources, we calculated the Poisson probability that the detected photons were a chance superposition of background photons and filtered the detected sources on this statistic. By summing this probability over our entire source list, we estimate that our catalogue contains $\sim$6 false sources, less than 1\% of the entire catalogue. The presence of associations at other wavelengths was used to identify faint X-ray sources, discarding sources without optical or near-IR counterparts. A Monte Carlo simulation was also used to confirm the authenticity of the faintest sources extracted using this method.

We employed a Bayesian technique to find optical or near-IR counterparts for our X-ray sources using deep, high-resolution optical and near-IR photometry from the IPHAS, 2MASS, and UKIDSS-GPS surveys. We found optical or near-IR associations for 1501 (89\%) of our sources and six band $r'$, H$\alpha$, $i'$, $J$, $H$, and $K_s$ photometry for 755 (45\%) of sources. This high success rate in finding counterparts for our X-ray sources is a testament to the recent generation of deep surveys covering the Galactic plane. Monte Carlo simulations were used to quantify the potential sources of error in this process, revealing that the highest source of error when cross-matching sources in such a crowded field is due to mis-associating X-ray sources with incorrect optical or near-IR counterparts.

We present the results of this work in three tables including X-ray observational data, properties of modeled spectra, broad-band X-ray fluxes, variability analysis results, and photometry in the six optical and near-IR bands. This catalogue represents one of the largest X-ray catalogue of a single SFR to date, the high density of sources in the two fields testament to the size of this massive SFR. This catalogue will serve as the foundation for future studies of Cygnus~OB2 and will be followed by a detailed analysis of the structure, dynamics, and star formation history of the region. The catalogue will also be useful for future studies of this region at other wavelengths and highlights the pivotal status of this region in terms of Galactic star formation.

\vspace{1cm}
\acknowledgments

We thank the anonymous referee for excellent suggestions that have greatly improved this work. 
We thank Patrick Broos (Penn State) for expert assistance with data analysis using the ACIS {\sc Extract} software. 

This research has made use of data from the {\it Chandra X-ray Observatory} (operated by the Smithsonian Astrophysical Observatory on behalf of NASA) obtained from the {\it Chandra} Data Archive. 
This publication makes use of data products from the INT Photometric H$\alpha$ Survey (IPHAS) carried out at the Isaac Newton Telescope (INT; operated on the island of La Palma by the Isaac Newton Group in the Spanish Observatorio del Roque de los Muchachos of the Instituto de Astrofisica de Canarias), the Two Micron All Sky Survey (2MASS; a joint project of the University of Massachusetts and the Infrared Processing and Analysis Center / California Institute of Technology, funded by NASA and the NSF), and the United Kingdom Infrared Telescope Deep Sky Survey (UKIDSS, supported by the UKATC and CASU). 
This research has made use of data obtained using the UK's AstroGrid Virtual Observatory Project, which is funded by the Science and Technology Facilites Council and through the EU's Framework 6 programme. 
This research has made use of NASA's Astrophysics Data System and the Simbad and Vizier databases, operated at CDS, Strasbourg, France. 

JJD was funded by NASA contract NAS8-39073 to the {\it Chandra X-ray Center} (CXC) during the course of this research and thanks the CXC director, Harvey Tananbaum, and the science team for advice and support. 
NJW acknowledges an SAO Pre-doctoral Fellowship.

\bibliography{/Users/nick/Documents/Work/tex_papers/bibliography.bib}

\clearpage

\clearpage

\begin{table}
\begin{center}
\caption{Log of observations} 
\label{log}
\begin{tabular}{ccccccc}
\tableline 
ObsID & Start Time & Exposure & R.A. (J2000.0) & Dec. (J2000.0) & Mode & Roll angle \\
	&		& (ks)	& (h:m:s)		& (d:m:s)	&	& (deg) \\
\tableline
4501 	& 2004 July 19 02:03:29 & 49.35 & 20:32:07 & +41:30:30 & {\sc vfaint} & 170.4 \\
4511 	& 2004 Jan 16 10:51:14 & 97.76 & 20:33:11 & +41:15:10	& {\sc vfaint} & 349.0 \\
\tableline 
\end{tabular} 
\end{center}
\end{table} 

\begin{table*}
\begin{center}
\caption{Results of the Monte Carlo simulation of the Bayesian cross-matching method}
\label{mcresults}
\begin{tabular}{llcc}
\tableline
&& Simulation 1 & Simulation 2 \\
\tableline 
Simulated sources$^*$	& Associated		& 0.85 	& 0.69	\\
					& Unassociated		& 0.15 	& 0.31	\\
\tableline
Associated sources$^\dagger$	& Correctly matched		& 0.93	& 0.91 	\\
					& Mis-matched				& 0.06	& 0.07	\\
					& Not matched				& 0.01	& 0.02	\\
\tableline
Unassociated sources$^\dagger$& Correctly unassociated	& 0.87	& 0.88	\\
							& Incorrectly associated	& 0.13	& 0.12	\\
\tableline
Observed sources$^\ddagger$	& Associated			& 0.86	& 0.72	\\
							& Unassociated		& 0.14	& 0.28	\\
\tableline 
\end{tabular} 
\end{center}
$^*$ These fractions are those put into the Monte Carlo simulation of the cross-matching.\newline
$^\dagger$ These fractions summarize the results of the Bayesian cross-matching scheme, highlighting which sources that were modeled as having true OIR counterparts were correctly matched to their counterpart or not.\newline
$^\ddagger$ These fractions summarize the fraction of associated and unassociated sources that would be observed following the results of the simulation.
\end{table*}

\begin{landscape}
\begin{table}
\scriptsize
\caption{ X-ray point sources detected in Cygnus OB2. Only the first 25 rows are shown. The complete table, containing the full catalogue is available in the electronic edition of the journal.} 
\label{table1b}
\begin{tabular}{rrrcrrccrccrrrccccc}
\tableline 
No. & \multicolumn{1}{c}{RA} & \multicolumn{1}{c}{Dec} & $\theta$ & \multicolumn{1}{c}{Area} &  $f_{PSF}$ & \multicolumn{1}{c}{Sig.} & log($P_{not}$) & \multicolumn{1}{c}{Exp.} & $\bar{E_x}$ & log($P_{KS}$) & \multicolumn{1}{c}{Cnts} & \multicolumn{2}{c}{$\delta_{Cnts}$} & \multicolumn{3}{c}{Count rates ($\times$$10^{-3}$ s$^{-1}$)} & Field & Flags\\ 
\cline{15-17}
  & \multicolumn{1}{c}{(J2000)} & \multicolumn{1}{c}{(J2000)} & (\arcm) & \multicolumn{1}{c}{(pixels)} & & \multicolumn{1}{c}{($\sigma$)} & & \multicolumn{1}{c}{(ks)} & (keV) & & \multicolumn{1}{c}{(net)} & Upper & Lower & \multicolumn{1}{c}{Full} &\multicolumn{1}{c}{Soft} & \multicolumn{1}{c}{Hard} & & \\ 
\tableline 
1 & 20:31:16.12 & 41:26:32.0 & 1.28 & 1081 & 0.89 & 1.9 & -2.64 & 47.62 & 2.20 & -0.70 & 10.87 & 5.6 & 4.5 & 0.228 & 0.084 & 0.145 & 2 &   \\ 
2 & 20:31:19.22 & 41:29:34.2 & 0.67 & 700 & 0.90 & 4.2 & -6.00 & 47.61 & 1.80 & -0.08 & 31.10 & 7.4 & 6.3 & 0.653 & 0.402 & 0.251 & 2 &   \\ 
3 & 20:31:20.64 & 41:31:51.1 & 0.44 & 654 & 0.90 & 7 & -6.00 & 47.62 & 1.99 & -0.01 & 68.44 & 9.8 & 8.7 & 1.437 & 0.728 & 0.709 & 2 &   \\ 
4 & 20:31:21.05 & 41:29:03.5 & 1.02 & 634 & 0.91 & 2.1 & -3.43 & 39.35 & 3.51 & -1.34 & 10.94 & 5.3 & 4.1 & 0.278 & 0.032 & 0.246 & 2 &   \\ 
5 & 20:31:22.02 & 41:28:39.7 & 0.93 & 615 & 0.91 & 2.4 & -4.66 & 47.54 & 1.47 & -0.09 & 13.29 & 5.5 & 4.4 & 0.280 & 0.189 & 0.090 & 2 &   \\ 
6 & 20:31:22.26 & 41:29:30.4 & 0.92 & 563 & 0.91 & 2.1 & -3.35 & 47.55 & 1.50 & -0.11 & 11.70 & 5.5 & 4.4 & 0.246 & 0.187 & 0.059 & 2 &   \\ 
7 & 20:31:23.16 & 41:31:50.4 & 0.69 & 535 & 0.90 & 3.4 & -6.00 & 47.57 & 1.72 & -2.55 & 22.29 & 6.5 & 5.4 & 0.468 & 0.299 & 0.169 & 2 &   \\ 
8 & 20:31:23.50 & 41:27:16.8 & 0.85 & 627 & 0.90 & 2.8 & -5.83 & 44.59 & 2.31 & -0.32 & 16.50 & 5.9 & 4.8 & 0.370 & 0.155 & 0.215 & 2 &   \\ 
9 & 20:31:23.52 & 41:29:49.1 & 0.30 & 477 & 0.90 & 9.3 & -6.00 & 47.53 & 1.42 & -0.44 & 109.20 & 11.8 & 10.7 & 2.298 & 1.796 & 0.501 & 2 &   \\ 
10 & 20:31:23.68 & 41:30:39.9 & 0.87 & 481 & 0.90 & 2.3 & -4.34 & 44.74 & 1.74 & -0.15 & 11.96 & 5.2 & 4.1 & 0.267 & 0.187 & 0.081 & 2 &   \\ 
11 & 20:31:25.84 & 41:28:37.4 & 0.81 & 394 & 0.89 & 2.3 & -4.69 & 47.46 & 1.58 & -0.52 & 11.76 & 5.1 & 4.0 & 0.248 & 0.164 & 0.084 & 2 &   \\ 
12 & 20:31:25.87 & 41:30:33.6 & 0.86 & 374 & 0.90 & 2 & -3.58 & 44.57 & 1.72 & -0.21 & 9.71 & 4.9 & 3.7 & 0.218 & 0.171 & 0.047 & 2 &   \\ 
13 & 20:31:27.37 & 41:23:57.3 & 1.49 & 964 & 0.90 & 1.4 & -1.77 & 28.86 & 1.38 & -0.59 & 7.01 & 4.9 & 3.8 & 0.243 & 0.202 & 0.041 & 2 &   \\ 
14 & 20:31:27.26 & 41:24:58.7 & 0.99 & 794 & 0.90 & 2.5 & -4.19 & 44.51 & 1.51 & -1.16 & 14.63 & 5.9 & 4.8 & 0.329 & 0.281 & 0.048 & 2 &   \\ 
15 & 20:31:27.70 & 41:29:17.4 & 0.53 & 328 & 0.90 & 3.8 & -6.00 & 47.44 & 1.48 & -0.67 & 23.77 & 6.3 & 5.2 & 0.501 & 0.361 & 0.140 & 2 &   \\ 
16 & 20:31:28.02 & 41:28:27.4 & 0.55 & 335 & 0.89 & 3.6 & -6.00 & 47.42 & 1.48 & -0.02 & 22.86 & 6.3 & 5.2 & 0.482 & 0.333 & 0.149 & 2 &   \\ 
17 & 20:31:28.39 & 41:34:09.9 & 0.87 & 465 & 0.90 & 2.3 & -4.61 & 47.52 & 0.91 & -0.45 & 11.69 & 5.1 & 4.0 & 0.246 & 0.196 & 0.050 & 2 &   \\ 
18 & 20:31:30.46 & 41:29:16.4 & 0.36 & 233 & 0.89 & 5.1 & -6.00 & 47.38 & 2.05 & -3.03 & 38.12 & 7.5 & 6.4 & 0.804 & 0.376 & 0.428 & 2 &   \\ 
19 & 20:31:30.56 & 41:28:17.2 & 0.64 & 285 & 0.90 & 2.6 & -6.00 & 47.36 & 1.69 & -0.65 & 13.64 & 5.2 & 4.1 & 0.288 & 0.223 & 0.065 & 2 &   \\ 
20 & 20:31:31.08 & 41:26:60.0 & 1.15 & 339 & 0.90 & 1 & -1.33 & 44.53 & 4.04 & -0.33 & 4.17 & 4.0 & 2.8 & 0.094 & 0.036 & 0.057 & 2 &   \\ 
21 & 20:31:31.65 & 41:32:40.7 & 0.72 & 273 & 0.90 & 2.2 & -4.83 & 41.79 & 2.64 & -0.16 & 10.76 & 4.9 & 3.7 & 0.257 & 0.089 & 0.168 & 2 &   \\ 
22 & 20:31:33.27 & 41:33:33.1 & 0.37 & 269 & 0.90 & 5.4 & -6.00 & 47.18 & 3.73 & -0.49 & 41.74 & 7.8 & 6.7 & 0.885 & 0.062 & 0.823 & 2 &   \\ 
23 & 20:31:33.69 & 41:26:50.8 & 0.71 & 290 & 0.90 & 2.3 & -5.02 & 43.71 & 1.96 & -0.34 & 11.46 & 5.0 & 3.8 & 0.262 & 0.174 & 0.089 & 2 &   \\ 
24 & 20:31:34.31 & 41:28:11.0 & 0.65 & 184 & 0.89 & 2 & -4.76 & 47.29 & 2.66 & -0.49 & 8.93 & 4.4 & 3.3 & 0.189 & 0.107 & 0.082 & 2 &   \\ 
25 & 20:31:34.80 & 41:29:33.2 & 0.39 & 148 & 0.90 & 3.5 & -6.00 & 47.30 & 3.34 & -4.00 & 20.40 & 5.8 & 4.7 & 0.431 & 0.050 & 0.381 & 2 &   \\ 
\tableline 
\end{tabular} 
\end{table} 
\scriptsize
Notes. Col. (1): X-ray source number used in all source tables in this paper and in future papers. Cols. (2)-(3): X-ray source position (calculated as the mean background subtracted event position). Col. (4): off-axis angle (arcmin). Col. (5): area of the source extraction region (in pixels). Col. (6): fraction of the point spread function within the extraction region. Col. (7): source detection significance. Col. (8): logarithm of the Poisson probability that the source is a chance coincidence of background events. Values below -6.0 are listed as -6.00. Col. (9): full exposure time for each source derived from the mono-energetic exposure maps for the combined observations. Col. (10): background corrected median energy of all source photons in the full (0.5-8 keV) band. Col. (11): logarithm of the Kolmogorov-Smirnov probability that the source is not variable. Col. (12): net counts in the full (0.5-8 keV) band. Cols. (13)-(14): upper and lower 1$\sigma$ errors on the number of net counts. Cols. (15)-(17): net count rates in the full (0.5-8 keV), soft (0.2-2 keV) and hard (2-8 keV) bands. Col. (18): source field number (1: central field, 2: north western field, 3: overlap region). Col. (19): warning flags (E: the source lies in or on a chip gap or at the edge of the chip, N: the source is very close to another source, such that $f_{PSF} < 0.85$).
\clearpage 
\end{landscape}

\begin{table}
\small
\begin{center}
\caption{Results of X-ray spectral fitting. Only the first 10 rows are shown. The complete table is available in the electronic edition of the journal.} 
\label{table2b}
\begin{tabular}{rrcccclcc}
\tableline 
No. & \multicolumn{1}{c}{Counts} & Model fit & log($N_H$) & \multicolumn{1}{c}{$kT$} & \multicolumn{3}{c}{X-ray fluxes (erg cm$^{-2}$ s$^{-1}$)} \\ 
\cline{6-8} 
 & \multicolumn{1}{c}{(net)} & & (cm$^{-2}$) & \multicolumn{1}{c}{(keV)} & \multicolumn{1}{c}{log $F$} & \multicolumn{1}{c}{log $F_s$} & \multicolumn{1}{c}{log $F_h$} \\ 
\tableline 
1 & 10.87 & - & 22.19$^{}_{}$ & -$^{}_{}$ & 30.07$^{+0.18}_{-0.24}$ & - & - \\ 
2 & 31.10 & 1T & 22.33$^{+0.39}_{-0.65}$ & 1.03$^{+5.39}_{-0.35}$ & 30.95$^{+1.48}_{-0.67}$ & 30.88 & 30.12 \\ 
3 & 68.44 & 1T & 22.40$^{+0.20}_{-0.33}$ & 1.46$^{+3.17}_{-0.99}$ & 31.26$^{+0.37}_{-0.41}$ & 31.12 & 30.69 \\ 
4 & 10.94 & - & 22.76$^{}_{}$ & -$^{}_{}$ & 30.16$^{+0.17}_{-0.21}$ & - & - \\ 
5 & 13.29 & - & 21.65$^{}_{}$ & -$^{}_{}$ & 30.16$^{+0.15}_{-0.17}$ & - & - \\ 
6 & 11.70 & - & 21.70$^{}_{}$ & -$^{}_{}$ & 30.11$^{+0.17}_{-0.21}$ & - & - \\ 
7 & 22.29 & 1T & 22.11$^{+0.18}_{-0.88}$ & 1.25$^{+6.58}_{-0.90}$ & 30.55$^{+0.34}_{-0.48}$ & 30.44 & 29.87 \\ 
8 & 16.50 & - & 22.24$^{}_{}$ & -$^{}_{}$ & 30.28$^{+0.13}_{-0.15}$ & - & - \\ 
9 & 109.20 & 1T & 21.59$^{+0.32}_{-0.47}$ & 1.67$^{+2.64}_{-1.18}$ & 30.94$^{+0.19}_{-0.15}$ & 30.78 & 30.44 \\ 
10 & 11.96 & - & 21.99$^{}_{}$ & -$^{}_{}$ & 30.14$^{+0.16}_{-0.18}$ & - & - \\ 
\tableline 
\end{tabular} 
\end{center} 
Notes. Col. (1): X-ray source number. Col. (2): net counts in the full (0.5-8 keV) band. Col. (3): X-ray spectral model fit type: single-temperature thermal plasma model (1T) or no model fit but quantities determined using methods described in Section \ref{s-spectra}. (-). Col. (4): hydrogen column density from model fit or from the median energy for unfit sources, with upper and lower 90\% confidence intervals$^*$. Col. (5): thermal plasma temperature of model fit with upper and lower 90\% confidence intervals$^*$. Col. (6): logarithm of the X-ray luminosity in the full (0.5-8 keV) band from model fit or derived from the number of net counts for unfit sources (assuming a distance of 1.45 kpc), with upper and lower 90\% confidence intervals$^*$. Cols. (7)-(8): logarithm of soft (0.5-2 keV) and hard (2-8 keV) band luminosities (assuming a distance of 1.45 kpc).\newline $^*$ Uncertainties are missing either when XSPEC was unable to compute them or where they are so large that the parameter is effectively unconstrained.
\end{table}

\begin{landscape}
\begin{table}
\small
\caption{Optical and near-IR counterparts for X-ray sources in Cygnus OB2. Only the first 10 rows are shown. The complete table, containing the full catalogue is available in the electronic edition of the journal.}
\label{table3b}
\begin{tabular}{rccc lll p{0.01cm} lllcc cl}
\tableline 
No. & RA & Dec & \multicolumn{1}{c}{Offset} & \multicolumn{3}{c}{Optical photometry} && \multicolumn{5}{c}{Near-IR photometry} & \multicolumn{1}{c}{AC2007} & \multicolumn{1}{c}{Spectral}\\ 
\cline{5-7} \cline{9-13}
  & & & \multicolumn{1}{c}{(\arcs)} & \multicolumn{1}{c}{$r'$} & \multicolumn{1}{c}{$i'$} & \multicolumn{1}{c}{H$\alpha$} && \multicolumn{1}{c}{$J$} & \multicolumn{1}{c}{$H$} & \multicolumn{1}{c}{$K_s$} & Origin & Flags & \multicolumn{1}{c}{source} & \multicolumn{1}{c}{type}\\ 
\tableline 
1 & 20:31:16.03 & 41:26:32.2 & 1.07 & $20.83\pm0.19$ & $18.73\pm0.06$ & $20.19\pm0.17$ && $15.39\pm0.05$ & $14.01\pm0.03$ & $13.44\pm0.04$ & MMM & AAA & - &  \\ 
2 & 20:31:19.32 & 41:29:32.3 & 2.22 & $16.67\pm0.01$ & $15.34\pm0.01$ & $16.23\pm0.01$ && $12.92\pm0.03$ & $12.13\pm0.03$ & $11.78\pm0.03$ & MMM & AAA & - &  \\ 
3 & 20:31:20.60 & 41:31:51.1 & 0.50 & $15.66\pm0.00$ & $14.68\pm0.00$ & $15.32\pm0.00$ && $12.75\pm0.03$ & $12.08\pm0.03$ & $11.77\pm0.02$ & MMM & AAA & - &  \\ 
4 & 20:31:21.15 & 41:29:04.1 & 1.29 & - & - & - && 16.89 & $15.61\pm0.12$ & $15.05\pm0.17$ & MMM & UBC & - &  \\ 
5 & 20:31:22.14 & 41:28:40.3 & 1.55 & $16.70\pm0.01$ & $15.60\pm0.01$ & $16.38\pm0.01$ && $13.47\pm0.03$ & $12.86\pm0.03$ & $12.57\pm0.03$ & MMM & AAA & - &  \\ 
6 & 20:31:22.28 & 41:29:29.9 & 0.49 & $20.73\pm0.18$ & $18.96\pm0.08$ & $20.12\pm0.20$ && $15.87\pm0.07$ & $14.53\pm0.05$ & $14.07\pm0.07$ & MMM & AAA & - &  \\ 
7 & 20:31:23.32 & 41:31:50.7 & 1.77 & $17.84\pm0.02$ & $16.68\pm0.02$ & $17.39\pm0.03$ && $14.64\pm0.03$ & $13.82\pm0.03$ & $13.60\pm0.05$ & MMM & AAA & - &  \\ 
8 & 20:31:23.47 & 41:27:15.1 & 1.74 & $20.41\pm0.13$ & $18.61\pm0.06$ & $19.73\pm0.14$ && 15.55 & $14.30\pm0.05$ & 13.98 & MMM & UAU & - &  \\ 
9 & 20:31:23.56 & 41:29:48.9 & 0.51 & $15.82\pm0.01$ & $14.97\pm0.01$ & $15.39\pm0.01$ && $13.46\pm0.03$ & $12.79\pm0.03$ & 12.68 & MMM & AAU & - &  \\ 
10 & 20:31:23.63 & 41:30:40.7 & 0.93 & - & - & - && $15.41\pm0.05$ & $14.34\pm0.05$ & $13.94\pm0.06$ & MMM & AAA & - &  \\ 
\tableline 
\end{tabular} 
\end{table} 
Notes. Col. (1): X-ray source number. Cols. (2)-(3): position of OIR source that X-ray source was matched to (blank if no association found). Col. (4): offset between X-ray and OIR source positions. Cols. (5)-(7): IPHAS $r'$, $i'$, and H$\alpha$ magnitudes with errors. Cols. (8)-(10): near-IR $J$, $H$, and $K_s$ magnitudes with errors. Col. (11): flag indicating origin of near-IR photometry for each band (M: 2MASS, U: UKIDSS). Col. (12): flag indicating photometric quality for 2MASS magnitudes where used. Col. (13) source number in the catalogue of Albacete Colombo et al. (2007). Col. (14): spectral types from the literature. References are: W73: Walborn (1973), MT91: Massey and Thompson (1991), H99: Herrero et al. (1999), H03: Hanson (2003), K07: Kiminki et al. (2007), N08: Negueruela et al. (2008). The numbers indicate the source numbers in Massey and Thompson (1991), or in Comeron et al. (2002) for sources where a letter prefixes the number.
\clearpage 
\end{landscape}

\end{document}